# Recent advances in high-throughput superconductivity research

Jie Yuan[1,2], Valentin Stanev[3], Chen Gao[4], Ichiro Takeuchi[3] and Kui Jin[1,2,5]

[1] Beijing National Laboratory for Condensed Matter Physics, Institute of Physics, Chinese Academy of Sciences, Beijing, 100190, China
[2] Key Laboratory of Vacuum Physics, School of Physical Sciences, University of Chinese Academy of Sciences, Beijing, 100049, China.
[3] Department of Materials Science and Engineering, University of Maryland, College Park, Maryland 20742, USA.
[4] Beijing Advanced Sciences and Innovation Center of Chinese Academy of Sciences, Beijing 101407 China
[5] Songshan Lake Materials Laboratory, Dongguan, Guangdong 523808, China

E-mail: yuanjie@iphy.ac.cn; takeuchi@umd.edu; kuijin@iphy.ac.cn

**Abstract**

Superconducting materials find applications in a rapidly growing number of technological areas, and searching for novel superconductors continues to be a major scientific task. However, the steady increase in the complexity of candidate materials presents a big challenge to the researchers in the field. In particular, conventional experimental methods are not well suited to efficiently search for candidates in compositional space exponentially growing with the number of elements; neither do they permit quick extraction of reliable multidimensional phase diagrams delineating the physical parameters that control superconductivity. New research paradigms that can boost the speed and the efficiency of superconducting materials research are urgently needed. High-throughput methods for rapid screening and optimization of materials have demonstrated their utility for accelerating research in bioinformatics and pharmaceutical industry, yet remain rare in quantum materials research. In this paper, we will briefly review the history of high-throughput research paradigm and then focus on some recent applications of this paradigm in superconductivity research. We consider the role these methods can play in all stages of materials development, including high-throughput computation, synthesis, characterization, and the emerging field of machine learning for materials. The high-throughput paradigm will undoubtedly become an indispensable tool of superconductivity research in the near future.

# 1. Introduction

Ever since Kammerlingh-Onnes in Leiden observed the disappearance of electrical resistance of mercury at very low temperatures in 1911, the phenomenon of superconductivity has been intensely studied. Yet, the exact nature of this effect remained unclear until the Bardeen-Cooper-Schrieffer (BCS) theory was developed in 1957 [1, 2]. Establishing the physical parameters behind the critical temperature ($T_c$) then became one of the main research topics in the field. Observation of an isotope effect clearly demonstrated that $T_c$ in most then-known-superconductors is linked to the coupling between the electrons and the lattice, with the strength of the electron-phonon (el-ph) interaction playing a key role. Increasing el-ph coupling enhances $T_c$ as long as the atoms in the crystal only vibrate near their equilibrium sites. But as the coupling strength crosses from a weak regime to an intermediate regime, the normal metallic state itself becomes unstable and a competing state such as a charge density wave may arise, causing structural changes. Although early analyses pointing to a ceiling of $T_c$ at about 30 ~ 40 K proved too simplistic, it quickly become clear that achieving significantly higher critical temperatures would require unusual conditions or materials [3]. (One example is the recent remarkable observation of high $T_c$ in hydrogen-rich materials under extreme pressure [4])

In 1986, the ceramic compound Ba-La-Cu-O was found to become superconducting at 35 K by Bednorz and Müller in Zurich [5]. This milestone work not only brought hope for superconductors with $T_c$ above 40 K, but also inspired other physicists to expand their search by exploring compounds with complex crystal structures and containing multiple elements. Soon, the limit of 40 K was crossed [6, 7]; critical temperatures above that of liquid nitrogen (77 K) were realized [8, 9], and the highest $T_c$ reached 138 K in $Hg_{0.8}Tl_{0.2}Ba_2Ca_2Cu_3O_{8.33}$ (at ambient pressure) [10]. While these discoveries inspired a generation of researchers, the mechanism of superconductivity in this class of materials (dubbed cuprates) is still a mystery.

Before the discovery of high-$T_c$ cuprate (HTSCs) materials, most known superconductors were composed of no more than two elements and were predominantly elemental superconductors, alloys, and intermetallics. As the superconducting transition temperature went beyond the 40 K mark, the superconducting compounds became more complex and containing more elements; for example, $YBa_2Cu_3O_{6+\delta}$, $HgBa_2Ca_2Cu_3O_{8+\delta}$ and $Hg_{0.8}Tl_{0.2}Ba_2Ca_2Cu_3O_{8.33}$ are made up of 4, 5 and 6 elements respectively. On the other hand, the recent discoveries of Fe-based [11, 12], Cr-based and Mn-based superconductors [13-15] demonstrated that even compounds with elements once thought strongly detrimental to superconductivity have to be considered. Obviously, the number of possible combinations grows exponentially as more and more elements are added in the candidate materials. The number of compounds can be easily estimated within the group of natural elements; the order of magnitudes for possible combinations are $10^3$, $10^5$, $10^7$ and $10^9$ for binary, ternary, quaternary, and penternary compounds respectively. In addition, many other methods to modulate superconductivity are known: ultra-high pressure [16, 17], ultra-thin film deposition [18], superlattice architecture [19] and ionic liquid/solid gating [20, 21] are a few examples. Third, some HTSC materials are extremely sensitive to physical and chemical parameters. For instance, one percent variation in cation content can turn a copper oxide superconductor into an insulator [22]. One will have to do a vast amount of synthesis to construct a reliable and detailed phase diagram as a function of cation and anion contents [23]. What is more, a complete phase diagram of HTSC must necessarily be multivariate, i.e. alongside cation and anion substitutions it must include synthesis conditions, pressure, magnetic field, and other variables. It is clearly not feasible to efficiently construct such a multidimensional phase diagram using the traditional one-material-at-a-time experimental methodology.

In general, there are two main challenges in the experimental study of superconductivity: i) searching for novel superconductors in the enormous space of candidate compounds comprised of more and more elements; ii) delineating the key physical parameters that control the superconductivity, by means of establishing a reliable multidimensional phase diagram. Conventional experimental methods are not well suited to address these problems, and new paradigms and tools are required to boost the efficiency of superconductivity research [24, 25].

In bioinformatics and the pharmaceutical industry, similar challenges arose before the discovery of HTSCs. There is a huge number of possible gene combinations and drug formulas, and it is not realistic to synthesize and test them one by one. The idea of doing many tests at the same time – the so-called high-throughput strategy – was rapidly developed and put into practice. Ever since, high-throughput methods have been a key driving force behind the development of modern biology and medicine. This methodology was also gradually adopted in condensed matter physics and materials science.

The general high-throughput materials research procedure has the following steps - sample synthesis, characterization, analysis of the data. Significant increase of the efficiency necessitates the acceleration of every step, as well as substantial coordination among them. The high-throughput approach for materials development is sketched in figure 1.

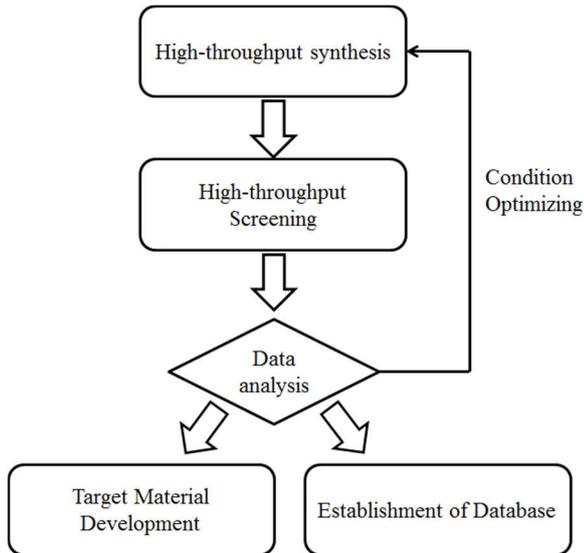

**Figure 1**. The high-throughput approach for discovery and development of materials.

In 1970, Hanak introduced the rudiments of a multiple-sample-concept in materials research, by developing a multiple-target, radio-frequency co-sputtering deposition [26]. However, the difficulties in obtaining and processing a large amount of experimental data prevented the wide-spread adoption of high-throughput methods in materials research at the time. Another factor inhibiting the spread of these methods was the low demand for complex functional materials. This has dramatically changed in the last decades, owing to the needs of industries such as communications, aerospace, energy, and transportations. The demand for novel materials has experienced an explosive growth, both in terms of diversity and required performance. This necessitated significant acceleration of materials research and development. High-throughput synthesis and characterization technology became a major research tool, leading to a substantial progress in a number of applications (see, for example, Refs. [27-34]). Helping these advances, the rapid progress in information technologies created powerful tools able to deal with the massive amounts of data generated by this approach.

Thus, high-throughput paradigm provides an extremely promising approach to meet the challenges in superconductivity research outlined above. In one early pioneering work demonstrating the potential of these methods, Xiang *et al.* reported the rapid synthesis of high-$T_c$ superconductors, with a 128-member library of copper oxide superconducting thin films deposited on a single substrate in one batch [35]. Although the superconductors synthesized in this work had been known already, the increase in efficiency achieved by the parallel synthesis was encouraging. In 2013, Jin *et al.* provided the first example of a superconductor discovered via the high-throughput methodology. They fabricated composition-spread films comprised of Fe and B, across 3-inch Si wafers by a co-sputtering technique.

Indications of a superconducting region were found from the resistance measurement by a home-made probe with an array of 64 pogo-pins (figure 2b). Superconductivity was then confirmed by a zoom-in magnetotransport and susceptibility measurements [36]. In 2016, Wu *et al.* studied a library of heterostructures made up of $La_2CuO_4$ and combinatorial $La_{2-x}Sr_xCuO_4$ ($0.15 \leqslant x \leqslant 0.47$), achieved by varying the deposition rate of the Sr content in an advanced oxide molecular beam epitaxy system [37]. In 2018, Stanev *et al.* filtered more than 100,000 compounds and created a list of potential superconductors, using a machine learning models trained on the critical temperatures of more than 12,000 known superconductors [38].

As seen from these examples, the high-throughput paradigm is already starting to permeate superconductivity research. In this review, we outline the advances in all parts of high-throughput superconductivity research: high-throughput synthesis (Section 2), high-throughput characterizations (Section 3), high-throughput *ab-initio* calculations (Section 4), and machine learning (Section 5). We also discuss the need to develop a next generation of facilities, techniques, and platforms in order to fully utilize the power of high-throughput methods (discussed in Section 6) [36].

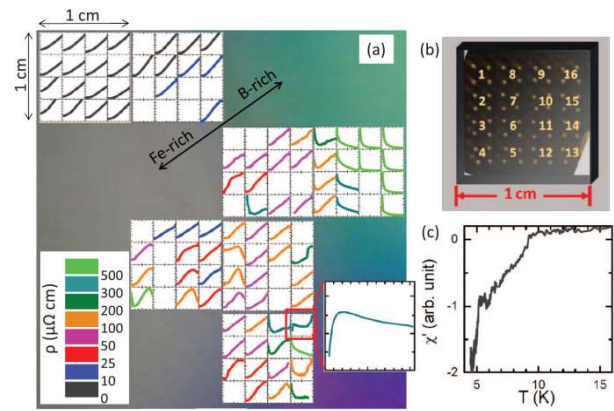

**Figure 2**. Mapping of the temperature dependence of resistivity on the Fe-B composition spread film.

## 2. High throughput syntheses for superconductors.

High-throughput synthesis is one of the foundations of the recent acceleration of the rate of materials exploration. However, that does not mean the synthesis cost has suddenly become irrelevant; efficiency is a key point – the speed of synthesis should grow faster than the resources it demands. To achieve this, common components and procedures used in different experimental steps, like evacuation, heating and conditioning atmosphere, should be combined as much as possible. This is the basis of the so-called combinatorial approach, which is the most efficient high-throughput synthesis method. Koinuma *et al.* gave the basic concept of combinatorial chemistry for solid state materials in Ref. [39]. During synthesis, a chemical reaction for the target material is influenced by many factors. The search for new materials can be regarded as a process of scanning certain points in the phase diagram as a function of multiple variables, as shown in figure 3. Compared to the classical synthesis process of an effective point-by-point search, combinatorial approach can much faster map key parameters (such as composition, temperature, and pressure) responsible for optimizing a functionality.

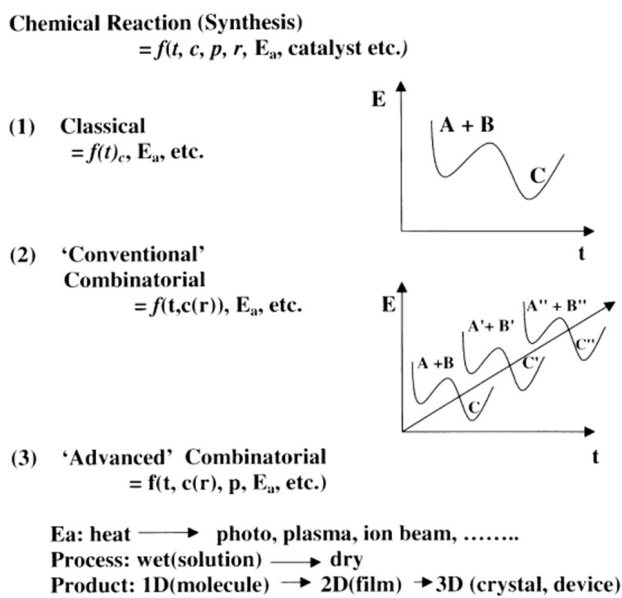

**Figure 3**. Concepts of combinatorial synthesis versus classical synthesis [39].

Thin-film preparation technology is the most widely used combinatorial synthesis method [30]. As early as 1965, Kennedy et al. reported a method for rapidly obtaining a Fe-Cr-Ni ternary library by electron beam co-evaporation of three precursors (Fe, Cr and Ni targets), on an equilateral-triangle metal foil with a side length of 10 inches [40]. Three generations of combinatorial film growth techniques have been developed since then. The first generation realizes a gradient chemical composition by ablating the precursors simultaneously and mixing them using natural diffusion. Both co-evaporation [40, 41], co-sputtering [26, 42, 43] and co-laser ablating [44] techniques are categorized into this generation. Figure 4 shows a typical configuration of a co-sputtering system. Different sources are mounted against the wafer at a certain angle to the normal direction of the wafer surface. During the deposition, such configuration results in a spatial variation of the deposition rate and allows plumes from different targets to overlap and form a natural composition gradient. Recently, Jin et al. obtained an Fe-B binary spread by such method, with a continuous composition across a 3-inch Si wafer with a 200 nm $SiO_2$ layer on top [36]. The lower part of figure 4 shows the photograph of one composition spread wafer taken under natural light; the average composition at different positions (solid circles) is obtained by the wavelength dispersive spectroscopy (WDS). A key advantage of the co-deposition method is that it requires relatively simple facilities and the deposition process is easy to control. However, the chemical composition cannot be precisely controlled because of the natural variability of the spatial deposition rate. The local composition of the film can only be determined by micro-region analysis methods (discussed in the section of high-throughput characterizations).

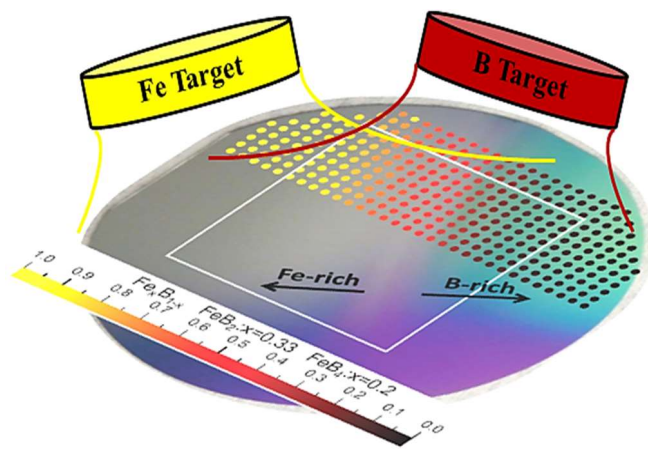

**Figure 4**. Sketch of the co-sputtering deposition for an Fe-B composition spread combi-film.

Second-generation methods were developed to overcome the shortcomings of the co-deposition approach. Xiang et al. used a method combining thin film deposition and physical mask techniques for a parallel synthesis of spatially addressable libraries of materials [35]. The programmable mask is at the core of the second-generation technology. Films fabricated with these methods contain discrete squares with varying compositions. Because these small regions are addressable, such films are also called integrated materials chips or combinatorial materials libraries. More details about this synthesis method can be found in the earlier literature (e.g. [32, 45-47]). It has to be noted that the mask is not necessarily a physical one. Wang et al. developed a combinatorial robot which is able to create thick-film libraries by ink-jet printing, as shown in figure 5 [48]. After being ground into sufficiently fine powder, precursors can be made into inks [49], taking the role of the three "primary colors". Different doses of inks are ejected on designated places of the wafer. Thus, using selected precursors and an appropriate post processing, an addressable materials library can be created. In this way, the ejector together with the robot arm replaces the physical mask [50, 51].

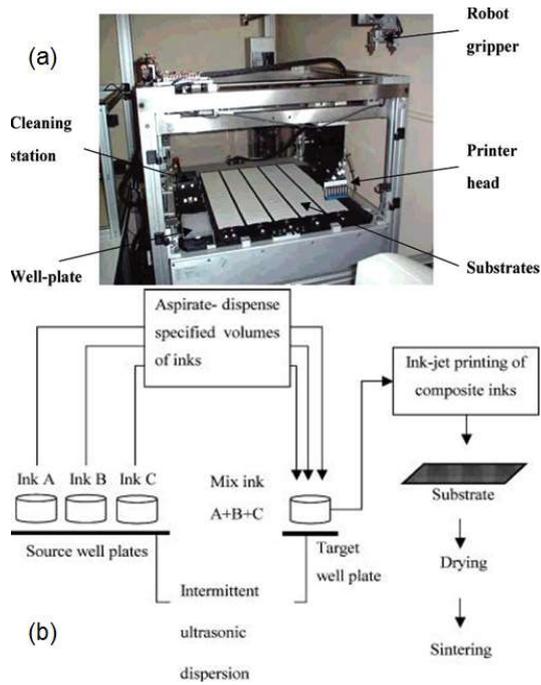

**Figure 5**. (a) The ink-jet printer table within the robot gantry, populated with alumina substrates. The pick and place robot arm is seen in the top right. (b) Schematic diagram of the mixing and printing protocol [48].

Mask patterns has been widely utilized in the study of electronic, magnetic, optical and dielectric materials, as well as catalysts and alloys [52]. However, for more precise studies of material properties, distinct techniques such as atomically controlled layer-by-layer thin-film growth are required [25].

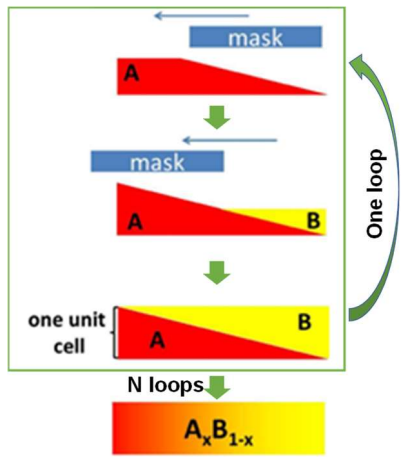

**Figure 6**. Schematic of binary-combinatorial-film growth using the continuous moving mask technique [53].

Combinatorial laser molecular beam epitaxy (CLMBE)–the so-called third generation of combinatorial thin-film preparation technology–was developed to carry out a parallel fabrication via a layer-by-layer growth mode [25]. The state-of-the-art laser molecular beam epitaxy in combination with a mobile mask technique has been already used to control layering sequences of high-quality superconducting films [54]. The procedure for preparing binary combinatorial films with continuous chemical composition spread on single substrate can be briefly described as follows (see also figure 6): once a target A is ablated by laser pulses, a metal mask is moving along the substrate in half a period time and results in a linear distribution of the A component. In the other half period, the other target B is ablated for a reverse distribution by moving the mask in the opposite direction [53]. A desired thickness of the combinatorial film can be achieved by setting corresponding periods of deposition. It should be emphasized that the two precursors A and B must be mixed in a single unit cell in the period, monitored by the reflection high-energy electron diffraction system for *in-situ* diagnostics. Otherwise, a superlattice rather than a combinatorial film will be obtained.

CLMBE is easy to operate, and benefits from a wide chemical composition range in one batch of deposition. This makes it perfect for obtaining a precise phase diagrams of materials like cuprate superconductors [5], For instance, Yu *et al*. fabricated combinatorial $La_{2-x}Ce_xCuO_{4\pm\delta}$ thin film with $x$ from 0.1 to 0.19 on $1\times 1$ cm$^2$ $SrTiO_3$ substrate by the CLMBE [53].

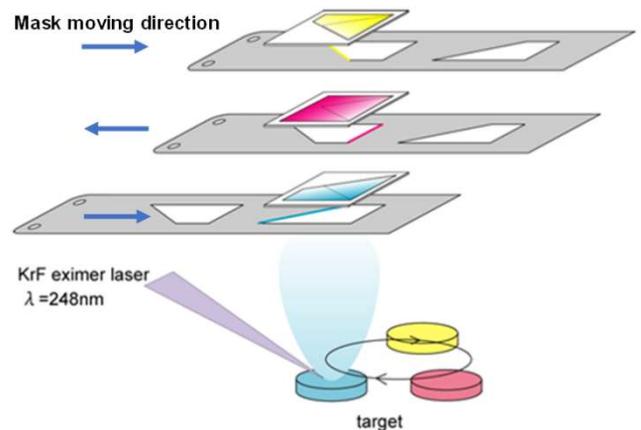

**Figure 7**. (From the introduction brochure of Pascal Co., Ltd in Japan) Schematic of ternary-combinatorial film growth using the continuous moving mask technique. The blue arrows indicate the direction of mask motion in every step.

Following the same principle, a ternary-combinatorial film can be grown with ablating alternatingly three targets and shadowing the substrate from three directions by mask with a sophisticated pattern design (shown in figure 7). With this method, Mao *et al*. prepared an Mg–Ni–Al ternary thin film library [24]. In combinatorial films, the chemical composition is locked to the fixed space point of the sample surface. For a ternary compound, the formula can be written as $A_xB_yC_{1-x-y}$, in which A, B and C stand for different elements. Since there are two independent spatial coordinators (e.g. $x$ and $y$) determining the position corresponding to a unique chemical composition, the full ternary library can be obtained based on single combinatorial spread. However, if more than two variables are involved, e.g. three variables in the quaternary $A_xB_yC_zD_{1-x-y-z}$, one cannot obtain all the combinations of these four components in a single chip; instead, a pseudo multi-component materials library can be created.

As explained above, three generations of distinct combinatorial-film preparation techniques have been developed. It has to be noted that each generation has its own unique merits, and is suitable for different kinds of materials. Sometime, a customized combinatorial approach is required, for example for compounds that only show superconductivity in a very narrow range of chemical doping. A typical iron-based superconductor, $FeSe_{1-\delta}$ only displays superconductivity within a finite $\delta$ [55]. Since Se is a volatile element, fine control of the Fe/Se ratio of the final film can be extremely challenging by the methods presented above. Recently, a double-beam laser configuration was used to produce a redistributed laser flux density on a FeSe target. As reported in the study of $SrTiO_3$ films grown by pulsed laser deposition (PLD), the stoichiometric ratio can be regulated by the power density of the laser [56]. Tuning the parameter of the combinatorial laser, the Fe/Se ratio of the final film can be varied by a very narrow δ step (i.e. the order of magnitude is ~ 0.001, which is beyond the resolution limit of the common chemical analysis methods). The left panel of figure 8 shows the schematic map of double-beam laser used in the deposition; the corresponding evolution of the superconducting critical temperature can be clearly seen in the normalized $R$-$T$ curves from different space regions of the final film (right panel). Moving along the film from one edge to the other, $T_c$ changes continuously from below 2 K to 12 K and back to below 2 K. There is no obvious difference in the film thickness (~150 nm) across the whole film (verified by scanning electron microscope (SEM)); thus, thickness influence on $T_c$ can be excluded. Such combinatorial films offer a good platform to investigate the nature of tunable superconductivity in the Fe-Se binary system [57].

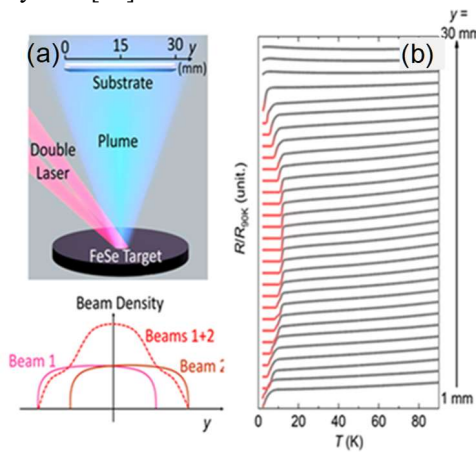

**Figure 8**. (a) Schematic process of the double-beam pulsed laser deposition via spectroscope, and then focused onto the target with controllable displacement. (b) The temperature dependence of normalized resistance along the $y$ direction, with the highest $T_c$ in the middle [57].

Wu *et al.* used a tilted Knudsen cell in an oxide molecular beam epitaxy system to synthesis combinatorial $La_{2-x}Sr_xCuO_4$ film. Although in principle this falls into the class of first generation combinatorial techniques, it provides a much finer control over the composition. The deposition rate can be precisely tuned and the growth process can be monitored in-situ. The composition gradient originates in the angle between the sources and the substrate, as shown in figure 9. By means of this method, the composition across the film can be confined to extremely fine steps, $\Delta x$ ~0.00008 [37, 58].

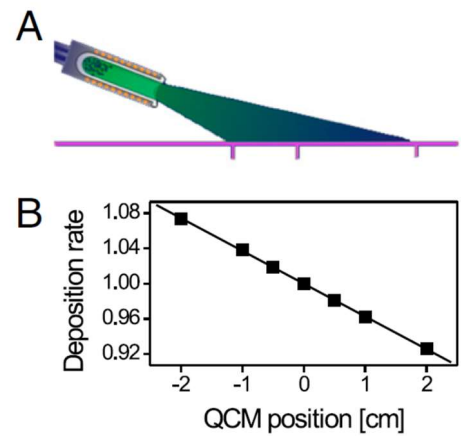

**Figure 9**. Schematics of the deposition geometry: thermal effusion cell is positioned at a shallow angle (20°) with respect to the substrate. Closer to the source, the deposition rate is higher [37].

In addition to generating composition-spread material libraries, combinatorial approaches can also be used to rapidly ascertain the optimal growth condition for superconducting materials. For example, the conventional way to find the best deposition temperature of a superconducting material is to try different temperatures one by one. However, a much more efficient method is to create a thermal gradient on the substrate during film deposition [59]. Figure 10 shows one configuration which can create such gradient during the growth process. One end of the substrate is attached to a heater while the other end is kept free-standing. When the heater is set at high temperature, heat diffuses to the free-standing end through the substrate. Using such configuration, a temperature range in which only pure *β*-FeSe phase forms was found; only three batches of samples were tested with deposition temperature from 350 to 700℃ [60].

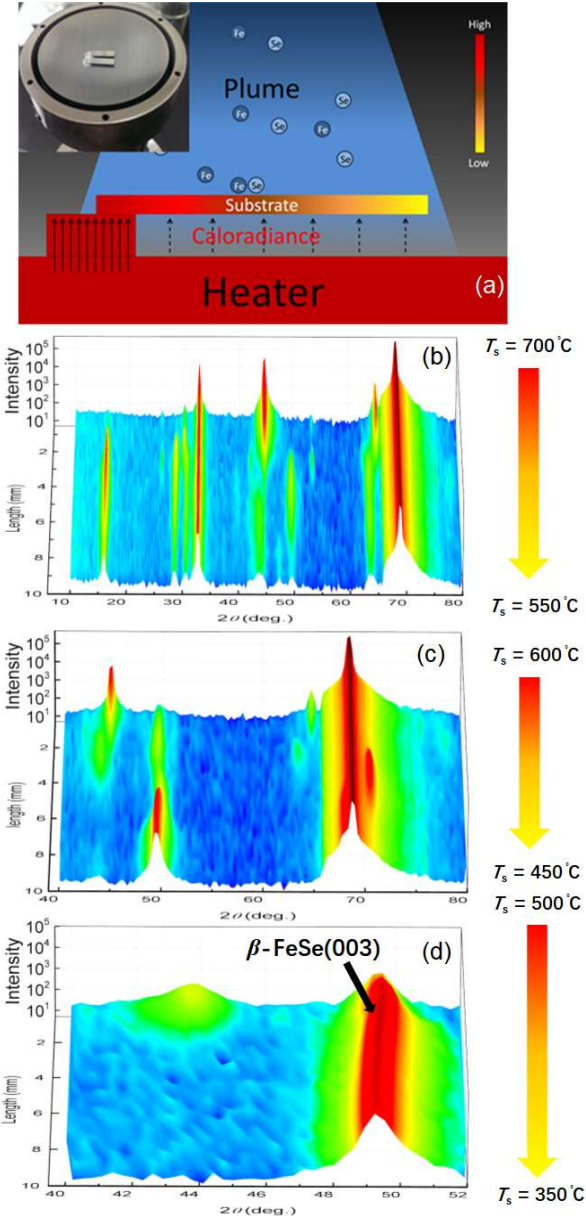

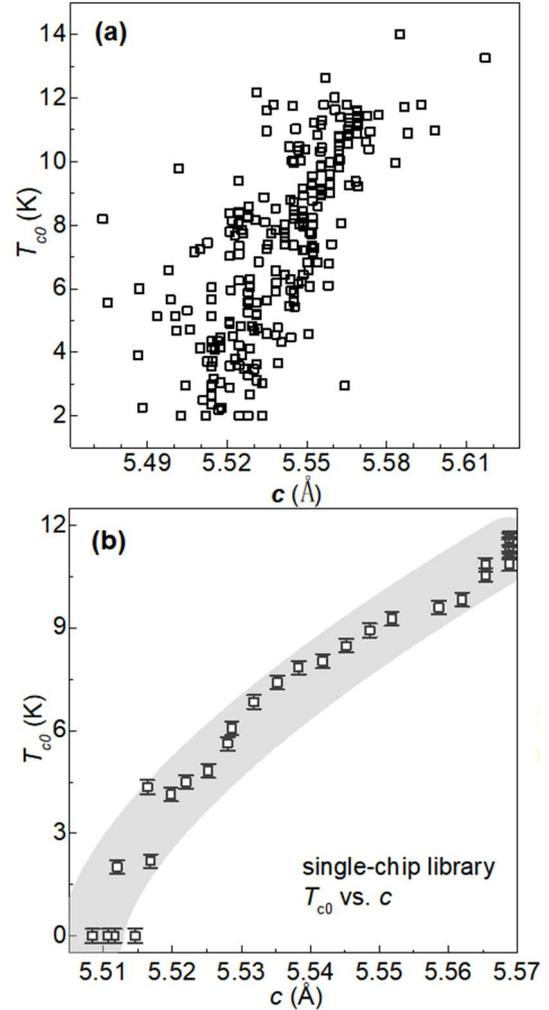

**Figure 10**. (a) Schematic of the combinatorial film growth with gradient temperature, inset on top left is real photo of the substrate mounted on the heater. (b-d) the micro-region $\theta/2\theta$ x-ray patterns of the FeSe films grown using different temperature gradients: 700 °C-550 °C, 600 °C-450 °C and 500 °C to 350 °C, respectively. Pure $\beta$-FeSe phase can be found in the region where the deposition temperature is not beyond 450 °C [60].

In addition, parallel synthesis can help reduce the uncertainty originating in the difference of growth conditions, unavoidable between separate experimental batches. Figure 11 shows the relationship between the out-of-plane crystal lattice parameter ($c$-axis) and $T_c$ extracted from more than one thousand pieces of uniform FeSe films (upper panel) as well as one combinatorial film (lower panel), respectively. It took almost three years to prepare and characterize the 1000+ uniform films; this is in contrast with the several weeks needed for the single combinatorial film. As can be seen, the amount of data collected from the uniform films is much larger, however, the positive relation between $c$ and $T_{c0}$ is manifested much clearly in the combinatorial film data.

**Figure 11**. The $c$-axis lattice constants dependence of $T_{c0}$ for (a) uniform FeSe films, (b) combinatorial FeSe film with gradient $T_c$ [57].

Combinatorial film data provides a better correlation between $T_c$ and lattice parameters due to its more precise composition control.

## 3. High throughput characterizations of superconductors

As another step in the high throughput methodology, rapid characterization of combinatorial films is indispensable in the drive to accelerate superconductivity research. Because to the nature of combinatorial synthesis, characterization have to be conducted by probes with spatial resolution capabilities. Several commercially available probes satisfy this requirement: atomic force microscope, scanning tunneling microscope (STM), optical microscope, scanning electron microscope are just a few examples. These tools with auto scan functionality can have high spatial resolution, and are equipped with tips or aggregated beams. There are also diverse homemade probes designed to study various properties of high-throughput samples. In this section, we focus on techniques relevant to superconductivity research, briefly summarized in the following order.

*3.1. Composition and structure*

Common probes used for composition and structure analyses are: x-ray diffraction; scanning electron microscopy (SEM); transmission electron microscopy (TEM). Because the electron beam can be focused, methods using it naturally provide the required spatial resolution and can be used for high-throughput characterizations [33, 61].

Great efforts have been dedicated to developing spatially resolved x-ray-based characterization methods. Since x-rays have much higher frequency and photon energy than visible light, they tend to penetrate or get absorbed in most materials. Unfortunately, most of the techniques used to redirect x-rays are unable to produce well focused beams.

Initially, pinholes or slits made of anti-radiation materials were used to shrink the profile size of x-ray beams. Figure 12 shows the data collected from a combinatorial $La_{2-x}Ce_xCuO_4$ library chip (an electron-doped copper oxide superconductor, $x$ varies from 0.10 to 0.19). Micro-area scan is realized by adding a narrow slot and a moving sample stage to a commercially-available x-ray diffractometer [53]. The calculated $c$-axis lattice constant monotonically decreases with raising the nominal doping level $x$, in accordance with the tendency extracted from uniform LCCO thin films fabricated by conventional pulsed laser deposition (PLD) method [62, 63].

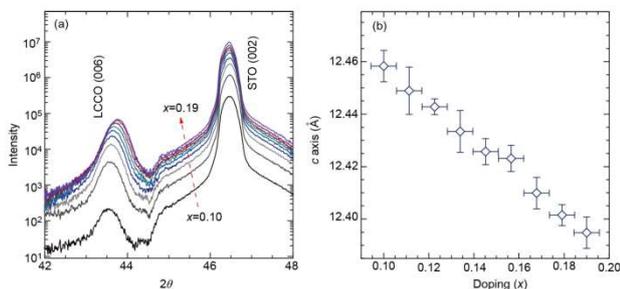

**Figure 12**. (a) The micro-region x-ray diffraction results of combi-film $La_{2-x}Ce_xCuO_{4\pm\delta}$. The component interval $\Delta x$ is about 0.012. (b) The variation of $c$-axis lattice constant with increasing doping levels. In this figure, $x$ refers to nominal doping level [53].

In this technique the analysis sensitivity and accuracy are unavoidably low, due to the limited incident x-ray throughput, which necessitates time-consuming repetitive angular-scanning measurements. To address this, techniques using a zone plate as an x-ray focusing component or a 2D detector were developed in order to enhance the intensity of the x-ray beam [64] or measurement efficiency [65]. In 2005, Luo *et al.* developed a combinatorial x-ray analysis system with a spot size of 0.5 mm by using polycapillary x-ray lens to focus the divergent beam from an x-ray source. This system, integrating micro-region x-ray diffraction, x-ray fluorescence, and x-ray photoluminescence tools, meets the requirements of high-throughput structure and composition characterizations. Instead of the angular-dispersive x-ray diffraction, these authors used energy-dispersive x-ray diffraction (EDXRD), which utilizes the entire spectra of the focused x-ray, from 4 to 19 keV, without any moving parts. As a consequence, the measurement process was further accelerated. The configuration used in the system is shown in figure 13 [66]. More details about the operation principle of x-ray optics can be obtained, for example, in the review paper Ref. [67].

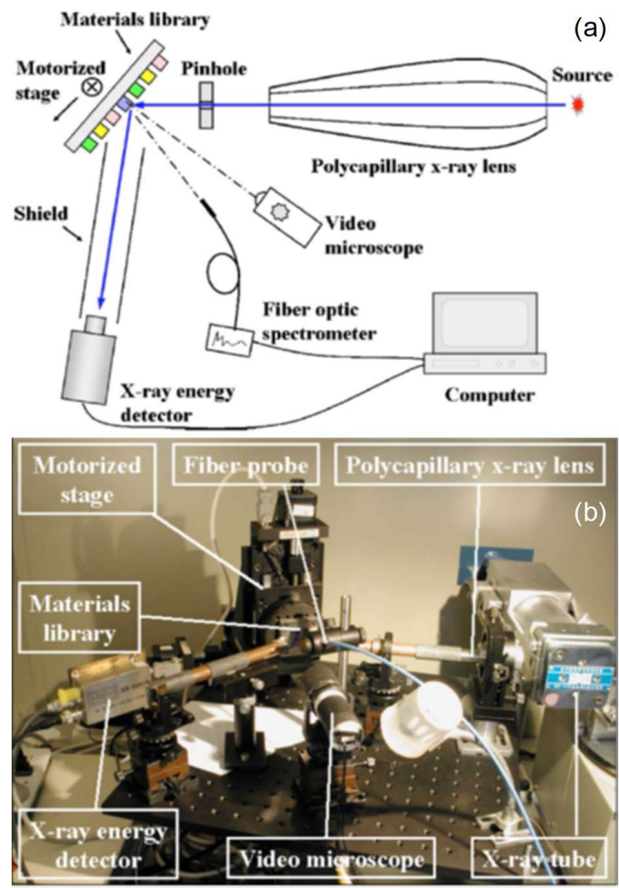

**Figure 13**. (a) Configuration and (b) photograph of the x-ray high-throughput characterization system [66].

Concurrently, special x-ray sources were developed to meet the requirements of high-throughput characterizations. Xiang *et al.* used a micro-beam x-ray system to simultaneously screen the composition and the structure with a resolution of 10-100 μm. The x-ray intensity in such system is 20 times higher than that of an ordinary x-ray source [32].

Since synchrotron facilities can offer high spatial resolution while simultaneously providing adequate photon flux, several high-throughput screening techniques were developed specifically for such sources. In 1998, Isaacs *et al.* used synchrotron x-ray micro-beam techniques to verify the feasibility of high-throughput characterization in combinatorial libraries [68]. This work included x-ray fluorescence, diffraction and absorption spectroscopy experiments with a spatial resolution of $3 \times 20$ μm$^2$. In 2014, combinatorial phase mapping techniques with a synchrotron source were realized by Gregoire *et al* [69]. The technique used in this work provides the capability of measuring more than 5000 samples per day using a synchrotron x-ray diffraction and fluorescence for rapid characterization. Xing *et al.* recently pushed the efficiency and automation even further [70]. The rate of recording diffraction images reached one pattern per second and the diffraction patterns were automatically processed to create the composition-phase map. In addition to commonly-used methods such as diffraction and fluorescence, x-ray absorption is also suitable for high-throughput characterization. For instance, Suram *et al.* used custom-made combinatorial x-ray absorption near edge spectroscopy technique to obtain a detailed chemical information of samples [71].

Although synchrotron x-ray facilities are powerful characterization tools, they remain a scarce resource, which impedes their use in high-throughput characterization. Developing more accessible rapid screening systems remains a grand challenge in the field.

*3.2. Transport properties*

Measuring transport properties such as resistivity, magnetoresistance and Hall effect is extremely important in uncovering the nature of superconductivity in various materials.

In 2005, Hewitt *et al.* developed a 196-pin device to measure the *dc* resistance versus temperature (shown in Figure 14). The distance between nearest pins is about 4.64 mm. In the configuration of Van der Pauw method, these pins can establish 49 parallel channels in an epoxy plate of 4.2 inches × 4.2 inches. For low-temperature measurement, the device can be cooled down to 7 K by cold fingers [72].

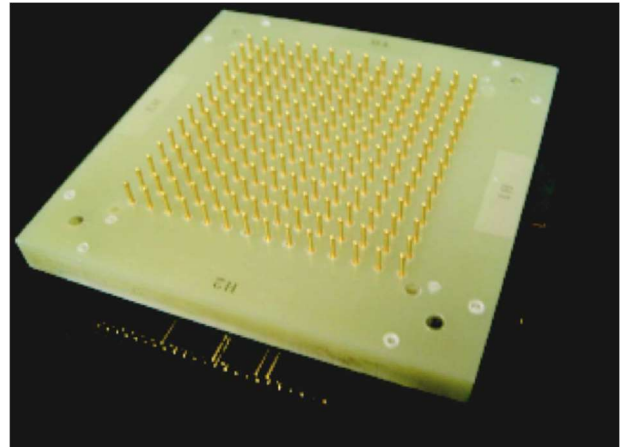

**Figure 14**. Image of the 196-pin device which makes electrical contact to the 49-sample thin-film library for four-contact resistivity measurements [72].

Typically, the sample chambers of most commercial low-temperature measurement system, especially the ones equipped with magnets, have limited space. For high-throughput magnetoresistance measurements, a more compact probe device, compatible with the existing systems, is highly desirable. In 2013, Jin *et al.* designed a high-throughput probe to search for superconductivity in the Fe-B composition spreads mentioned above. They used pogo pins for a better contact between the sample surface and the pinpoint, and also shortened the distance between nearest pins to ~1 mm. Thus, even if the lateral size of the sample chamber is limited to a centimeter scale, at least 16 multiplexed channels of temperature dependent resistance can be measured at $4 \times 4$ evenly spaced 1 mm × 1 mm areas in van der Pauw configuration. Utilizing this probe, the authors discovered superconductivity in the region where the ratio of Fe to B is approximately 1:2; there the superconducting transition temperature can reach 4 K. The temperature-dependent resistivity curves and a sketch diagram of the pogo pin array are shown in figure 2. A number of diced 1 cm × 1 cm chips from the superconducting region were measured using the 16-spot simultaneous resistance-versus-temperature measurement setup. Later, the device was upgraded and made compatible with a commercial physical property measurement system (PPMS). In this way, magnetoresistance and Hall data of a combinatorial film can be rapidly collected.

The group of I. Bozovic performed multiple-channel measurements of the Hall signal and the resistivity by a high-throughput system in order to characterize highly conductive oxides with Hall coefficients as small as $10^{-10}$ m$^3$/C. The receptacles of this device are shown in figure 15(a). There are 64 slots matching pogo pins (with 1.11 mm distance between the adjacent pins), which contact with the patterned film as shown in figure 15(b). In the center of the film, a stripe with 300 μm in width and 10 mm in length is bridged with 64 square pads covered with gold. The blocks for different channels segment the main stripe to 300 μm in length. Using this device, 30 channels for the resistivity and 31 channels for the Hall resistivity can be measured simultaneously. Figure 15 c) shows a series of longitudinal resistivity $\rho(T)$ curves from a single $La_{2-x}Sr_xCuO_4$–$La_2CuO_4$ bilayer films. The doping resolution can reach $\Delta x = 0.0002$ [73]. Due to the high special resolution of the Hall measurements, an intrinsic inhomogeneity of sample could be detected. Wu *et al.* were thereby able to construct a detailed phase diagram of this material and discovered a quantum charge-cluster glass state that competes with the superconducting state [37].

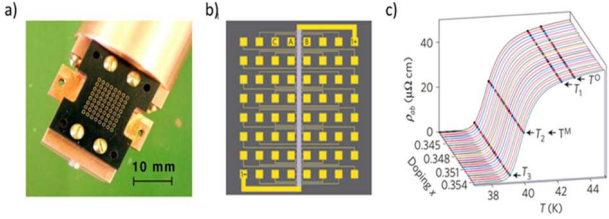

**Figure 15**. (a) Receptacles for spring-loaded pogo pins on the cryostat tail. (b) Film lithography and contact pattern. (c) The temperature-dependent *ab* plane resistivity $\rho$ measured from different channels on the same $La_{2-x}Sr_xCuO_4$-$La_2CuO_4$ bilayer films [37].

Since in this design the current bridge is shared by all the channels, the sample should have good electrical conductivity. Thus, this pattern is not suitable for screening a combinatorial film with insulating areas. In such case, an array micro-bridge design with independent current paths should be employed. In 2017, Wu *et al.* reported a 'sunbeam' lithography pattern for simultaneous Hall measurements [74]. The 36 Hall bars, each with three pairs of transverse contacts (as sketched in figure 16(a), form a circle with 10° between two successive bars. Angular dependence of $\rho_T$ (transverse resistivity, defined as $V_T d/I$, where $V_T$ is the voltage across the sample, transverse to the electrical current $I$ and $d$ is the film thickness) and $\rho$ in an under-doped ($x = 0.04$) LSCO film shows $\sin(2\varphi)$ oscillations; these suggest a breaking of the four-fold rotational symmetry of the crystal structure as shown in figure 16(b). The fact that this anisotropy emerges in superconducting samples is significant, as its origin was attributed to electronic nematicity.

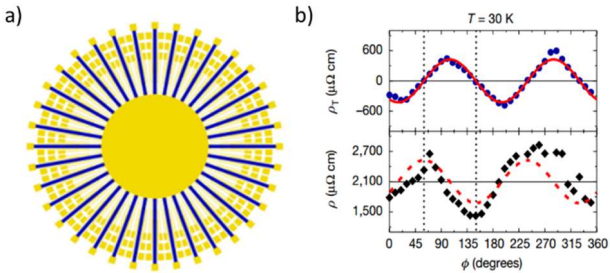

**Figure 16**. (a) The 'sunbeam' lithography pattern used for device fabrication contains 36 Hall bars, each with three pairs of transverse contacts. (b) the measured values of $\rho_T(\varphi)$ and $\rho(\varphi)$ at $T = 30$ K [74].

The spread of such methods is restricted by the difficulties of micro-fabrication. Shan *et al.* followed an alternative approach. They developed a scanning point-contact probe compatible with PPMS. By mounting the sample on a moving stage consisting of a close-loop z-axis and x-axis piezo cubes as shown in the left panel of figure 17, a Pt-Ir tip can be positioned on the sample surface with an accuracy of sub-micron scale. With this probe, He *et al.* investigated the tunneling spectra of [111]-, [110]-, and [001]-oriented high quality $LiTi_2O_4$ (LTO) thin films and unveiled anisotropic electron-phonon coupling in LTO system [75]. The ability to probe local physical properties of superconductors will allow to rapidly screen superconducting combinatorial films without the need for complex micro-fabrication.

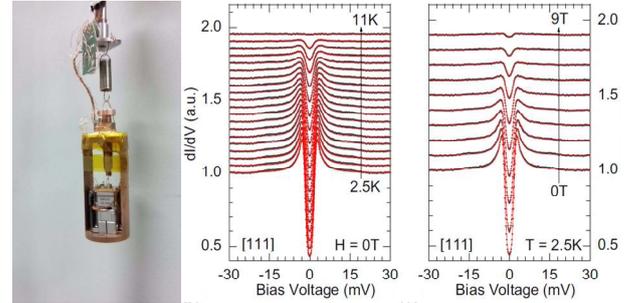

**Figure 17**. On the left: photograph of the scanning point-contact probe. Middle and right panels: normalized differential conductance versus bias voltage for the [111] film from 2.5 to 10 K and from 0 to 9 T, respectively [75].

### 3.3. Magnetic properties

Since common probes with beams focused by optical lenses are commercially available, here we skip the description of magnetic-optical methods for high-throughput characterization. Scanning magnetic microscopes, e.g. magnetic force microscope (MFM) [76-78], scanning Hall probe microscope (SHPM) [79, 80] and scanning superconducting quantum interference device (SQUID) [81-84], which possess the spatial resolution necessary for high-throughput magnetic imaging, have played and will continue to play an important role in understanding the nature of superconductivity.

MFM, with spatial resolution of about tens of nm, is generally based on measuring the force between a magnetized tip and the scanned surface. This method shows potential for high-throughput mapping in both static and dynamic magnetic fields [85]. MFM can study the topography of a sample and even image a single quantized vortex in a superconducting material [86]. SHPM provides quantitative magnetic field measurements with a spatial resolution better than 1μm under general conditions. A sub-micron Hall probe can be manufactured and used to scan a sample with the aid of conventional scanning techniques [87]. Similar to MFM, SHPM can also image the distribution of quantized vortices in type-II superconductors [88]. Local magnetic fields can also be imaged by scanning with SQUIDs, which utilize the effects of Cooper pair tunneling and superconducting quantum interference [89]. In contrast with MFM and SHPM, scanning SQUIDs can obtain information about the local susceptibility and superfluid density, as shown in figure 18 [90]. Due to the lack of suitable combinatorial samples, these techniques are still rarely applied in high-throughput characterization of superconductors. However, their advantages and enormous potential cannot be neglected. Kogan and Kirtley have modeled the Meissner response of anisotropic superconductors and superconductors with inhomogeneous penetration depths [91, 92]. This makes it possible to extract the evolution of the penetration depth from a local diamagnetic susceptibility imaging of combinatorial superconducting films.

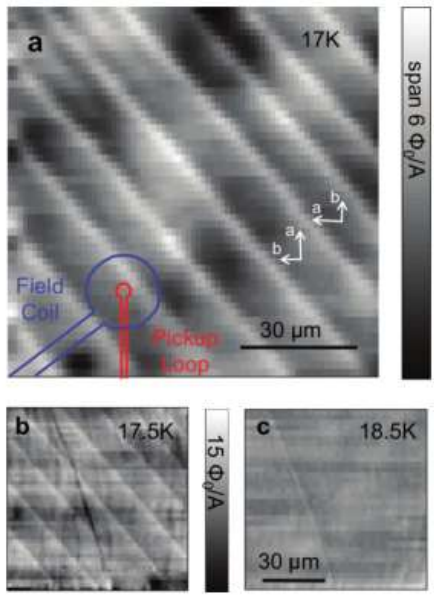

**Figure 18**. Local susceptibility image of the *ab* face of underdoped Ba(Fe$_{1-x}$Co$_x$)$_2$As$_2$ with *x*=0.051 and $T_c$=18.25 K. (a) at *T*=17 K, sample, (b) at *T* = 17.5 K, (c) at *T* =18.5 K. The brighter patterns indicate the regions with stronger diamagnetic response, which disappear above $T_c$ [90].

In addition to the transport and magnetic properties characterization tools mentioned above, near-field detection techniques like scanning tip microwave near-field microscope (STMNM) was developed to probe the dynamical electromagnetic properties of superconductors.

Wei *et al.* used a sharpened solid metal (instead of an aperture or a gap) as a point-like evanescent field emitter. Based on this design, they set up a near-field rf/microwave microscope that is capable of imaging surface resistance profiles with a spatial resolution of ~5 μm [93]. Later, Gao *et al.* improved the spatial resolution to the order of 100 nm by introducing a sapphire disk with a center hole comparable to the diameter of the tip wire and a metal coating layer of 1 μm on the surface, as shown in figure 19. In addition, a phase-sensitive technique was applied for faster data acquisition, while quasi-static theoretical model was developed to calculate the relative dielectric constant from the raw data [94].

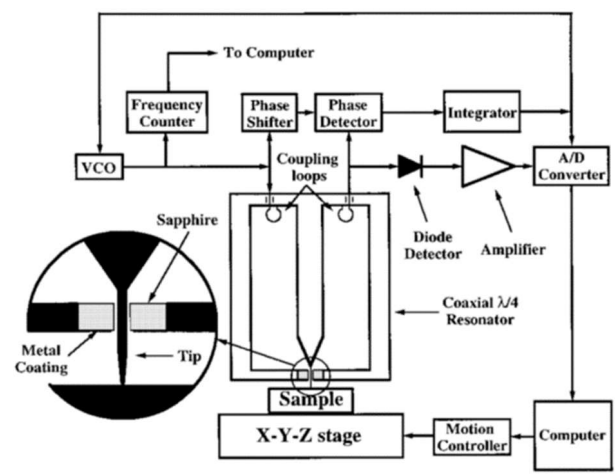

**Figure 19**. Schematic of the experimental setup for the STMNM with a resonator [93].

In the same year, Takeuchi *et al.* integrated the STMNM with a low-temperature cryostat and applied it to scan an YBa$_2$Cu$_3$O$_{7-\delta}$ film. The high-frequency response was used to detect the superconducting transitions at different positions on the film [95].

Shen group designed a near-field scanning microwave microscope in a different way. They separated the sensing probe from the excitation electrode to suppress the common-mode signal and used co-planar waveguides to transmit microwave signal (shown in figure 20) [96]. Similarly, their microscope was later upgraded for low-temperature imaging [97].

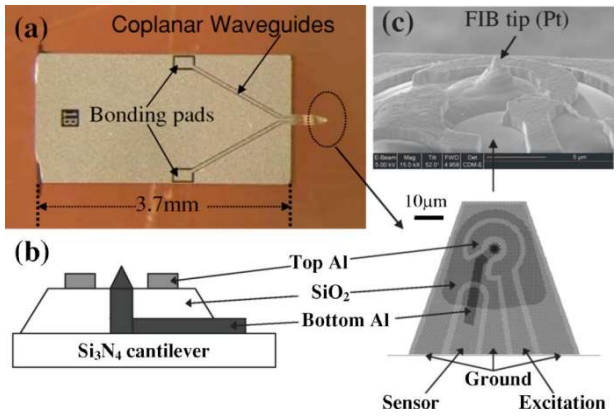

**Figure 20.** (a) Microfabricated cantilever with co-planar waveguides patterned on the top. The bonding pads and the chip dimension are labeled in the picture. (b) Schematics of the side view (left) and the top view (right) of the cantilever tip. All electrodes are labeled in the figure. (c) Scanning electron microscope (SEM) image of the Pt tip formed by focused ion beam (FIB) deposition [96].

In addition to composition, structure, electrical transport, and magnetism, methods were developed to image thermal, optical, and mechanical properties of films [98-103]. From these measurements some key information about superconductivity can be obtained.

For example, Otani *et al.* set up an integrated scanning-spring-probe system for high-throughput thermoelectric power-factor screening [101]. The schematic diagram can be seen in figure 21. Thermal transport measurements such as Nernst and thermopower are complementary to electrical transport results, and can be indispensable in clarifying issues related to multiband features [104], superconducting fluctuations [105] and phase transitions [106].

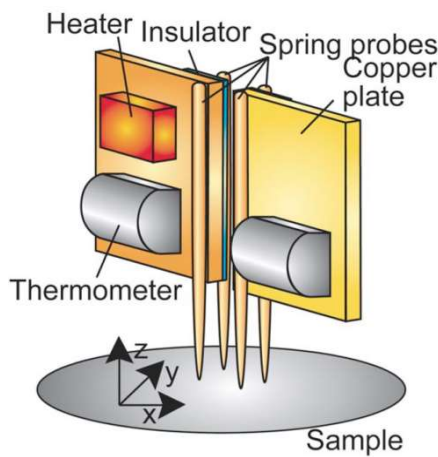

**Figure 21.** Schematic diagram of the probe to measure thermoelectric property [101].

An imaging ellipsometer with a lateral resolution in the nanometer range was invented by Karageorgiev *et al.* As shown in figure 22, the system combines an ellipsometer with atomic force microscope (AFM). The sample is illuminated from the bottom through a prism in total internal reflection (TIR), and the evanescent field created at the sample surface is scattered by a conventional AFM tip [107, 108]. The intensity of the measured optical signal depends on the sample thickness and local optical properties. By fitting the data and comparing with an existing database, the local thickness and other parameters such as refractive index and absorption factor can be extracted [109, 110]. This can be extremely useful for studying some interesting topics like a possible negative index of refraction induced by surface Josephson-plasma waves in layered superconductors [111].

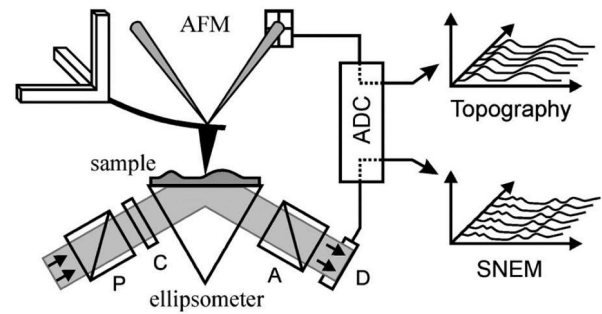

**Figure 22.** Experimental setup of scanning near-field ellipsometric microscope [107].

## 4. High throughput theoretical calculations for superconductors

*4.1. Introduction to high-throughput computations*

In combinatorial materials studies, the space of possible combinations is so enormous that even high-throughput experiments can rarely or never cover it completely. So, in order to achieve acceptable scale and rate of screening and optimizing, the number of candidate systems has to be restricted with the help of some heuristic or theoretical reasoning [112-115]. By using high-throughput calculations based on *ab-initio* methods, molecular dynamics and Monte Carlo simulations, etc., variety of properties for multiple materials can be predicted at a high rate, and serve as starting points for subsequent experiments. Furthermore, the combination of high-throughput computational results and experimental data can facilitate the construction of materials databases. These, in turn, can be mined and analyzed using materials informatics and artificial intelligence tools. Exploring the relations among the components, structure, and properties of materials can provide ideas for new materials designs.

High-throughput computations have already been employed in some research fields, such as solar water splitters [116, 117], solar photovoltaics [118], topological insulators [119], scintillators [120, 121], $CO_2$ capture materials [122], piezoelectrics [123], thermoelectrics [124, 125], hydrogen storage materials [126, 127], Li-ion batteries [128-133].

*4.2. High-throughput computations for superconductors*

There have been several notable successful examples of discovery of superconductors with the help of high-throughput computations. Kolmogorov *et al* undertook an intensive study of the Fe-B system via *ab-initio* high-throughput evolutionary calculations [134]. The crystal structure and formation energy for the entire Fe-B system was calculated, as shown in figure 23. Calculation of the electron-phonon coupling predicted $oP$10-FeB$_4$ to be a conventional superconductor. Stimulated by this theoretical study, a high-throughput combinatorial research was subsequently devised and executed (details are given above) [36].

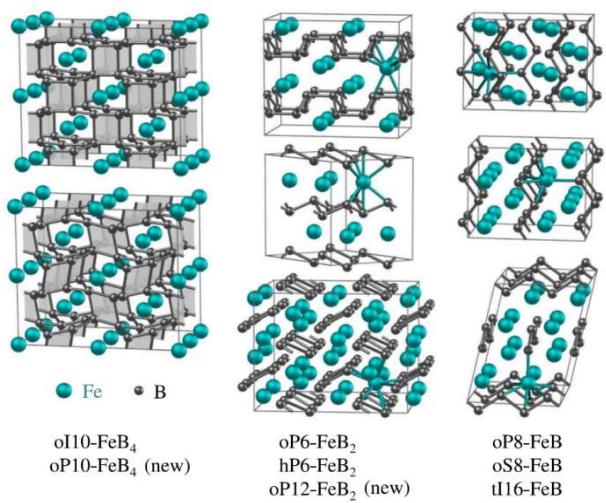

**Figure 23**. Competing B-rich Fe-B phases [134].

Matsumoto *et al* employed high-throughput computations to search for new thermoelectric or superconducting materials [135, 136]. By data-driven materials search, they extracted from a database of inorganic materials (AtomWork) thousands of candidates materials with some specific characteristics (e.g. flat band). Screening further the first-principles calculations, SnBi$_2$Se$_4$ and PbBi$_2$Te$_4$ were selected from all these candidates on the basis of their electronic structure (gap size, density of states near the Fermi level, etc.). The authors successfully synthesized SnBi$_2$Se$_4$ single crystal via melt and slow-cooling method (the exact composition is SnBi$_{1.64}$Se$_{3.53}$ showed by Energy Dispersive X-Ray Spectroscopy). Superconducting transition was indeed observed under pressure. A sister compound PbBi$_2$Te$_4$ also shows pressure-induced superconductivity, as shown in figure 24.

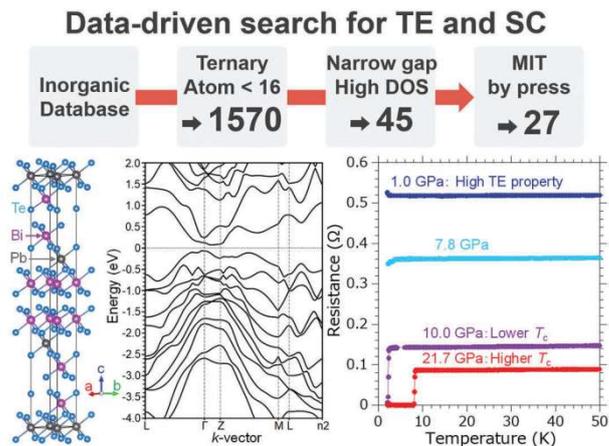

**Figure 24**. (a) Crystal structure, (b) band structure, (c) temperature dependence of resistance around superconducting transitions in PbBi$_2$Te$_4$ under various pressures [136].

Other studies have applied *ab-initio* high-throughput calculations on much larger scale. Klintenberg and Eriksson defined several distinct characteristics of the band structure of cuprates as key ingredients of their superconductivity [137]. They used these characteristics as criteria to filter through the calculated electronic structures of all materials in the Inorganic Crystallographic Structure Database (ICSD). Over 100 materials were identified as potential HTSC. Similar approach was used by Geilhufe *et al.*, with p-terphenyl–an organic material that shows sign of a superconducting transition at $T_c \approx 123$ K when doped with potassium–used as a prototype [138]. A database containing the electronic structures of more than 10,000 organic crystals was searched for compounds with similar density of states (DOS). As a result, 15 materials were proposed as candidate HTSC.

In the case of conventional superconductors, the mechanism of superconductivity is well understood, so their superconducting properties can be calculated from first principles, at least in theory. However, the specific mechanisms behind most examples of unconventional superconductivity are still not clear, and the electrons in these materials are often strongly correlated. Thus, attempts to predict new unconventional superconductors through deductive calculations are impossible even in principle. Nevertheless, high-throughput computations can at least provide some rough guidance in the search for new superconductors.

High-throughput computations also speed up the construction of database containing basic properties of new materials, which can be combined with machine learning methods to predict new superconductors (see next section), thus providing a new pathway for superconductivity research [138].

## 5. Machine learning for new superconductors

High throughput experiments often result in large data sets. Since it can be slow and impractical to analyse this data manually, the analysis step has become a common bottleneck for the entire process. Recent developments, however, allow a different approach to data analysis and, even more generally, to exploration of superconductors and the factors that determine the critical temperatures of various materials. Machine learning and statistical methods can use the information amassed in databases covering various measured and calculated materials properties in order to predict macroscopic variables, circumventing to need for theoretical models and complicated *ab-initio* tools (for more details see, for example, Seko *et al.* [139]).

The difficulties of modelling strongly correlated materials such as HTSCs provided an early impetus to the search for novel approaches. Application of statistical methods in the context of superconductivity began with simple clustering methods [140, 141]. These studies demonstrated that three "golden" descriptors confine the sixty known at the time superconductors with $T_c > 10$ K to three small islands, while other superconductors are randomly dispersed in this space. Based on this observation, predictions for potential high-temperature superconductors were made. Another early work by Hirsch used statistical methods to look for correlations between normal state properties and $T_c$ of the metallic elements in the first six rows of the periodic table [142].

These studies remained isolated early attempts to use data-driven approaches for study of superconductivity. However, the recent significant increase of available data made possible the widespread use of machine learning algorithms to address a multitude of research questions. An investigation by Isayev *et al.* in 2015 again found a clustering property of HTSC materials using structural and electronic properties data as predictors [143]. A classification model separating superconductors into two groups according to their $T_c$ was developed and showed good performance. As a demonstration of the impact that machine learning can have in the search for superconductors, a sequential learning framework designed by Ling *et al.* to discover the material with the highest $T_c$ was evaluated on ~ 600 known superconductors [144]. The framework did significantly better than a method based on random guessing.

A work by Stanev *et al.* in 2018 built upon many of the lessons from the earlier studies [38]. Whereas previous investigations explored several hundred compounds at most, this work considered more than ten thousand different compositions. These were extracted from the SuperCon database, created and maintained by the National Institute for Materials Science in Japan. The database contains information such as $T_c$ for an exhaustive list of all reported superconductors, as well as related non-superconducting compounds, including many closely-related materials varying only by small changes in stoichiometry. This unique information permitted to study crucial subtleties in chemical composition among related compounds. The order-of-magnitude increase in training data also allowed to optimize and automate the machine learning model construction procedures.

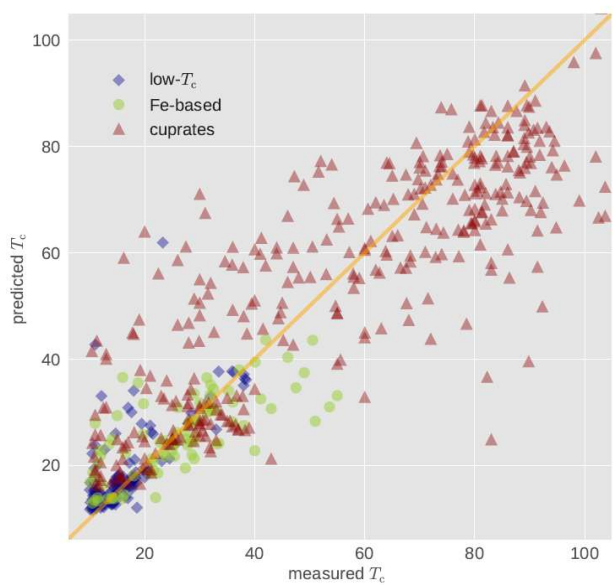

**Figure 25**. Performance of the regression model for $T_c$ presented in Ref. [38]. Notice that there is a clear correlation between measured and predicted values.

From SuperCon, a list of approximately 16,400 compounds was extracted, of which 4,000 have no $T_c$ reported. Of these, roughly 5,700 compounds are cuprates and 1,500 are iron-based (about 35% and 9%, respectively), reflecting the significant research efforts invested in these two HTSC families.

To analyse the data, first a classification model designed to separate materials into two distinct groups depending on whether $T_c$ is above or below some predetermined value was developed. The temperature that separates the two groups - $T_{sep}$ - was treated as an adjustable parameter of the model. For $T_{sep} = 10$ K, the proportion of above-$T_{sep}$ materials is approximately 38%, while model accuracy is about 92%, i.e., the model can correctly classify nine out of ten materials, which is much better than random guessing.

Several regression models designed to predict the actual $T_c$ values were presented, focused on materials with $T_c > 10$ K (see figure 25). Separate models were constructed and trained on materials from cuprate and iron-based HTSC families, as well as with all other materials denoted "low-$T_c$" for brevity. Differences in the used predictors across the family-specific models reflect the fact that distinct mechanisms are responsible for driving superconductivity in these groups.

Finally, the classification and regression models were integrated in a pipeline and employed for a high-throughput screening of the entire ICSD database for candidate superconductors. First, a classification model was applied on this dataset to create a list of materials with predicted $T_c$ above 10 K. The list of candidates contained about 3,000 materials. Next in the pipeline, the list was fed into a regression model trained on the entire SuperCon database to predict $T_c$. Filtering for the materials with $T_c > 20$ K, the list was reduced to about 2,000 compounds. The vast majority of the compounds identified as candidate superconductors were cuprate HTSC materials. There are also some materials clearly related to the iron-based superconductors. The remaining set has 35 members, and is composed of materials that are not obviously connected to any known superconducting family. The list comprises of several distinct groups; especially interesting are the compounds containing heavy metals (such as Au, Ir, Ru), metalloids (Se, Te), and heavier post-transition metals (Bi, Tl), which are or could be pushed into interesting/unstable oxidation states.

Although at the moment it is not clear if any of the predicted compounds are superconducting, the presence of one highly unusual electronic structure feature in many of the candidates is encouraging; almost all of them exhibit one or several flat (or nearly flat) bands just below the energy of the highest occupied electronic state. Such bands lead to a large peak in the density of states and can cause a significant enhancement in $T_c$. Attempts to synthesize several of these compounds are already underway.

In another very recent work, Xie *at al.* utilized machine learning to obtain a closed-form approximation for $T_c$ of phonon-driven superconductors, using an automatic algorithm to test a number of different functional forms [145]. The formula selected by the algorithm is more accurate than the commonly used approximation due to McMillan, Allen and Dynes, demonstrating once again the great promise of machine learning in this field.

## 6. Conclusion and perspective

In this short review, some recent developments in superconductivity-related high-throughput techniques are summarized. Revealing the mechanisms of high-temperature superconductivity and discovering new unconventional superconductors have always been important topics in condensed matter physics. As materials under study become more and more complex, the advantages of high-throughput methodology become more pronounced. Although these advantages have proven to be significant, some limitations should also be pointed out. Currently, most high-throughput experiments are based on film techniques. This brings some specific constrains. Effects like lattice mismatch and surface aging have to be considered. The number of possible combinations are limited, and the use of in-situ measurements is often required. Moreover, in view of the reduced dimensions and surface reconstruction effects, the experimental results obtained from film samples may not reflect the intrinsic properties of bulk samples [146-149]. Such issues are present not only for superconductors, but for many other materials as well. In order to further accelerate the research on superconductors, as well as other functional materials, advances in the following three aspects are highly desired.

### 6.1. New high-throughput techniques

While many combinatorial film preparation techniques have been developed, there is a great demand for high-precision characterization instruments, especially ones that can perform advanced spectroscopy measurements such as STM and angle resolved photoemission spectroscopy (ARPES). This requires combinatorial material libraries in the form of films to be measured *in-situ*. Accordingly, combined systems such as combinatorial-film-STM and combinatorial-film-ARPES tools will be optimal. These will be especially beneficial for the study of superconducting combinatorial libraries. Characterization methods taking advantage of synchrotron sources should become widespread; these techniques allow the study of not only composition and structure of materials, but also their elementary excitations. For instance, resonant inelastic x-ray scattering (RIXS) technique can be used to measure spin excitation [150], charge excitation [151], phase of the order parameter [152], etc. Importantly, thin films can be probed with RIXS, since it only needs small sample volumes. Compared with the neutron scattering techniques, RIXS appears much more promising in high-throughput research on superconductivity[153].

### 6.2. Future high throughput superconductivity research style

The high-throughput materials research paradigm is quite different from the traditional approaches, which sometimes can be compared to a "needle-in-a-haystack" search. Hanak proposed an integrated materials development workflow about 40 years ago [26]. Potyrailo and Mirsky extended the workflow by adding new elements, such as planning of experiments and data mining, as shown in figure 26 [154].

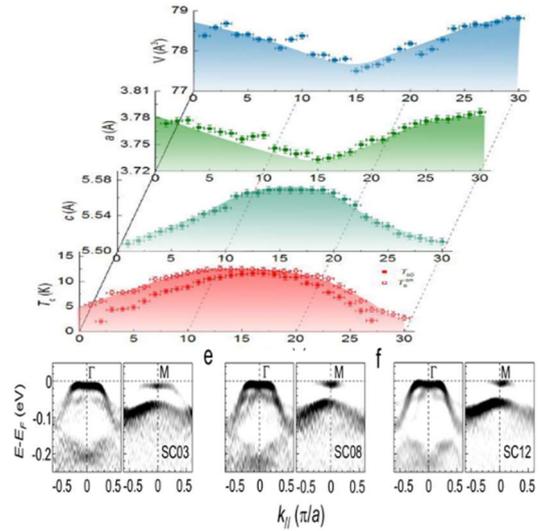

**Figure 27**. For FeSe film: the evolution of lattice constants from the $T_c$ gradient combinatorial in the upper panel; the band structure of three uniform samples with different $T_c$ obtained from ARPES in the lower panel [57].

### 6.3. MGI platform

High-throughput methods with their enormous potential to accelerate materials research have already attracted significant attention. Some countries have started projects and programs in order to support the development of high-throughput computational tools. The US Department of Energy created Scientific Discovery through Advanced Computing (SciDAC) in 2001 to develop new computational methods for tackling some of the most challenging scientific problems. In 2008, The US National Academies created a new discipline, Integrated Computational Materials Engineering, aimed to design materials using a variety of software tools either simultaneously or consecutively. In 2011, the US launched the Materials Project, a core program of the Materials Genome Initiative, that uses high-throughput computing to uncover the properties of all known inorganic materials. "Advanced Concepts in *ab*-initio Simulations of Materials" project, proposed by European Science Foundation, is aimed to develop rapid *ab-initio* calculations, which allow parameter-free calculations of real materials at the atomic level and are applicable to condensed matter systems.

However, in order to maximize the efficiency and impact of high-throughput methodology, it is necessary to closely coordinate the theoretical and experimental stages and combine computations with experiments and analysis of materials databases. One possible way of doing this is to integrate all the facilities in a big platform and organize them as nodes in an assembly line. The collected data can be standardized to establish a sharable databank. Techniques such as STM, neutron scattering and ARPES can contribute to

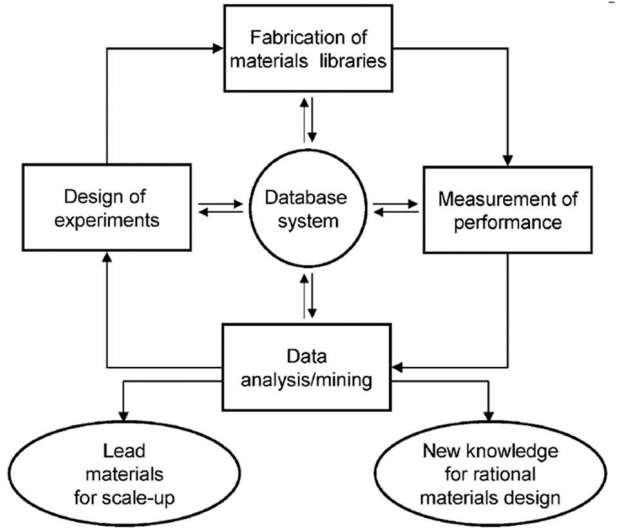

**Figure 26**. Typical combinatorial materials development cycle [154].

Superconductivity research will certainly benefit from more widespread adoption of this paradigm, but it has its own specifics which should be taken into account. Currently, some characterization methods such as STM, neutron scattering and ARPES, commonly used in the field, lag significantly behind the requirements of high-throughput screening. Thus, research on superconductivity, especially on mechanisms of high-temperature superconductivity, will have to combine studies of combinatorial libraries with exploration of conventional uniform films or bulk crystal samples. This was exemplified in the recent work on the relation between micro-structure and $T_c$ in FeSe superconductor. In this study the evolution of electrical transport properties and crystal lattice parameters were delineated by high-throughput experiments, whereas the electronic band structure was obtained from ARPES on uniform films (see figure 27) [57].

the data collection according to their limitations and characteristics. Since the MGI project was first announced in the US in 2011 [155], several similar programs have been initiated in different countries. Accelerated Metallurgy (AccMet) project initiated by the European Union and "MatNavi" database managed by National Institute for Materials Science in Japan are programs established to develop novel materials research and developments procedures. Among these programs, the Center for Materials Genome Initiative (CMGI) in Beijing, China may be the only one currently organized as a specialized high-throughput material research platform. This platform contains three large substations shown in figure 28: high-throughput computation and database station, high-throughput synthesis and fast characterization station, and high-throughput technique research and development station. MGI and similar platforms have the potentials to revolutionize materials research in the near future.

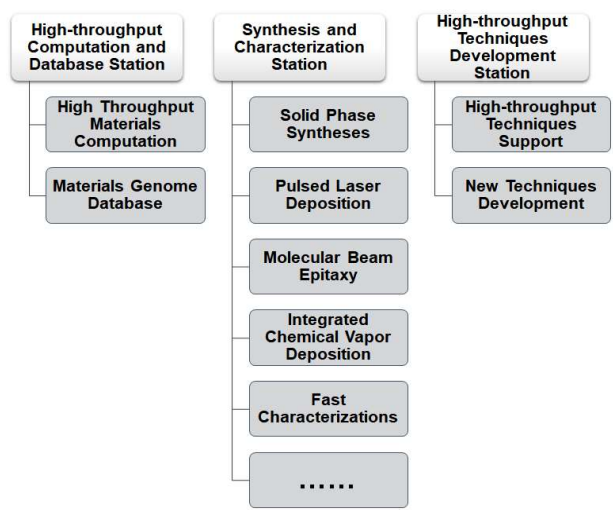

**Figure 28.** The organization of Center for Materials Genome Initiative, Huairou, Beijing.

In conclusion, integrating a novel research paradigm with new techniques, facilities, and platforms, high-throughput methodology is becoming more and more prominent in advancing the study of superconducting materials.

**Acknowledgments**

The authors would like to thank M Y Qin, X Zhang, X Y Jiang, G He, Z P Feng, D Li, R Z Zhang, and J S Zhang for the help during the manuscript preparation. This work was supported by the CAS Interdisciplinary Innovation Team, the Strategic Priority Research Program of Chinese Academy of Sciences (XDB25000000), the National Key Basic Research Program of China (2015CB921000, 2016YFA0300301, 2017YFA0303003, and 2017YFA0302902), the National Natural Science Foundation of China (11927808, 11674374, 11804378 and 51672307), and the Key Research Program of Frontier Sciences, CAS (QYZDJ-SSW-SLH001 and QYZDB-SSW-SLH008). The work at the University of Maryland was supported by ONR (N000141512222 and N00014-13-1-0635), and AFOSR (FA 9550-14-10332).